\documentclass[11pt, a4paper]{article}

\usepackage[a4paper, total={16cm, 25cm}]{geometry}

\usepackage{subfigure}% Support for small, `sub' figures and tables. Your choice of alternative may be preferred instead.
\usepackage{natbib}
\usepackage{multirow}
\usepackage{url}
\usepackage{amsmath}
\usepackage{graphicx}

\begin{document}

\LARGE
\begin{center}
  \textbf{Statistical models for short and long term forecasts of snow depth} \\[5mm]
  Hugo Lewi Hammer\footnote{Email: \texttt{hugo.hammer@oslomet.no}}
\end{center}
    
\large

\begin{center}
%  Hugo Lewi Hammer\footnote{Email: \texttt{hugo.hammer@oslomet.no}}\\
  OsloMet -- Oslo Metropolitan University\\
  Norway
\end{center}

\normalsize

\begin{abstract}
Forecasting of future snow depths is useful for many applications like road safety, winter sport activities, avalanche risk assessment and hydrology. Motivated by the lack of statistical forecasts models for snow depth, in this paper we present a set of models to fill this gap. First, we present a model to do short term forecasts when we assume that reliable weather forecasts of air temperature and precipitation are available. The covariates are included nonlinearly into the model following basic physical principles of snowfall, snow aging and melting. Due to the large set of observations with snow depth equal to zero, we use a zero-inflated gamma regression model, which is commonly used to similar applications like precipitation. We also do long term forecasts of snow depth and much further than traditional weather forecasts for temperature and precipitation. The long-term forecasts are based on fitting models to historic time series of precipitation, temperature and snow depth. We fit the models to data from three locations in Norway with different climatic properties. Forecasting five days into the future, the results showed that, given reliable weather forecasts of temperature and precipitation, the forecast errors in absolute value was between 3 and 7 cm for different locations in Norway. Forecasting three weeks into the future, the forecast errors were between 7 and 16 cm.
\end{abstract}

\section{Introduction}

The amount of snow, or snow depth, is important to many applications like road safety \citep{Juga13}, risk assessments \citep{Bocchiola06,Blanchet11}, winter sport activities \citep{sf,fnugg}, hydrology \citep{Jonas09} and climate change research \citep{Brown03,McCabe10,Falarz02,Scherrer13,Zhang04}.

The modeling of snow depth is typically divided into three parts: snow accumulation (snowfall), snow aging and melting. The physics behind the different parts is quite complicated and depends on many factors. For snowfall, a common rule is the 10:1 rule stating that the density of the arriving snow is one tenth of the density of water. Following this rule, 10 mm of precipitation results in 10 cm of snow. In reality, the relation is more complicated. The density of snowfall is related to the ice-crystal
structure by virtue of the relative proportion of the occupied volume of crystal composed of air. Snow density
is regulated by in-cloud processes that affect the
shape and size of growing ice crystals, subcloud
processes that modify the ice crystal as it falls, and ground-level compaction due to prevailing weather conditions and snowpack metamorphism. Understanding
how these processes affect snow density is difficult because direct observations of cloud microphysical processes, thermodynamic profiles, and surface measurements are often unavailable. \citet{Roebber03} builds a neural network to classify snow density of snowfall to three classes heavy (1:1 $<$ ratio $<$ 9:1), average (9:1 $<$ ratio $<$ 15:1), and light (ratio $>$ 15:1) where ratio refers to the density of water compared to the density of the arriving snow. The authors use the predictors solar radiation, temperature, humidity, precipitation and wind. The method only classify to correct density class in 60\% of the cases which emphasize the difficulty in forecasting snow density and snow depth. Snow aging and melting of snow and resulting changes in snow depth depends on many factors like temperature, precipitation, solar radiation and the age and density of the snow pack. Compared to other factors, precipitation and temperature are the main drivers of changes in snow depth and most snow models are only based on these factors \citep{Brown03,Kohler06}. 

Weather forecast services routinely forecast quantities like temperature, precipitation, wind and air pressure, but very rarely snow depth or other snow related quantities. An exception is \verb|snow-forecast.com| \citep{sf} which forecast snow depths for skiing resorts around the world, but the methods used are not publicly available.

The small number of weather services forecasting snow depth or other snow related quantities are in a big contrast to the amount logged snow depth data available around the world. E.g. in the US and Canada daily snow depth are logged for at least 8000 locations \citep{Brown03}. Motivated by the lack of models to forecast future snow depths and the large amount of historic snow depth data available, in this paper we present models to fill this gap. We build both short-term forecasts, where reliable forecasts of temperature and precipitation are available, and forecasts going further into the future. The work in \citet{Brown03} is related to the work in this paper, except that the authors use a numerical model to relate snow depth to precipitation and temperature, while we present a statistical model such that probabilistic forecasts can be performed.

The paper is organized as follows: In Sections \ref{sec:short} and \ref{sec:long} we describe the statistical models for short and long-term forecasts of snow depth. The models are analyzed on real data in Section \ref{sec:example} and the paper ends with some closing remarks in Section \ref{sec:closrem}.

\section{Short term forecasting of snow depth}
\label{sec:short}

Temperature and precipitation are the main drivers of changes in accumulated snow depth. Better forecasts of future snow depth can therefore be achieved by including weather forecasts of temperature and precipitation in the forecast model. Such a model will be presented in this section.

Let $D_t$ denote the current snow depth at a specific time of day at day $t \in \{1,2,\ldots,n\}$ where $n$ is the number of days with observations. Further let $R_t$ and $T_t$ denote the total precipitation and average air temperature for the last 24 hours. 

Snow depth will for a large portion of days be zero (no snow), and else always larger than zero. Precipitation has the same properties and is typically modeled by a zero-inflated gamma model \citep{Stern84,Sloughter07,Moller13}. The model results in a good fit also to the snow depth data and can be formulated as follows
\begin{equation}
  \label{eq:5}
  f(D_t) = P(D_t = 0)\, \mathrm{I}(D_t = 0) + P(D_t > 0) g(D_t)\, \mathrm{I}(D_t > 0)
\end{equation}
where $g(D_t)$ is the gamma density and $\mathrm{I}(x)$ return $1$ if $x$ is true and 0 else.

We model $g(D_t)$ as a generalized linear model, and the expectation of the gamma density is linked to covariates as follows. We follow the typical assumption of changes in snow depth by dividing in snowfall (increase) and snow aging/snow melt (decrease) \citep{Kohler06,Kuusisto84,Brown03}. First we find a model for snowfall. If the temperature is sufficiently cold, precipitation will arrive as snow, and if the temperature is sufficiently warm, the precipitation will arrive as rain. For temperatures around zero $^{\circ}$C, precipitation will arrive as a mixture of snow and rain. Several functions are suggested to model this, see e.g. \citet{Wen13,Kienzle08}. We use the inverse logit function which fits well to our data and is also in accordance with observations from earlier research, see e.g. Figure 6 in \citet{Kienzle08}. More specifically we assume that the expected depth of snow from $R_t$ mm of precipitation at temperature $T_t$ $^{\circ}$C is given by
\begin{equation}
  \label{eq:1}
  R_t\, \beta_0\, \mathrm{logit}^{-1}(\beta_1 + \beta_2 T_t)
\end{equation}
where 
\begin{equation}
  \label{eq:2}
  \mathrm{logit}^{-1}(x) = \frac{e^x}{1+e^x}
\end{equation}
The parameter $\beta_0$ refers to the snow water density ratio and is normally set to 10 \citep{Roebber03}. We will estimate the parameter from snow depth observation. We expect that $\beta_2$ will be negative such that higher temperatures result less snow (lower snow depth). 

The snowpack tend to sink with time, but less if it is very cold. Further, for temperatures around and above zero $^{\circ}$C the snow will additionally melt (transformed to water). The aging/melting is going faster if it is raining on the snowpack and is typically modeled with the index method (see e.g. \citet{Scherrer13}) which simply is an interaction between the temperature and the amount of rain, $(\beta_4 + \beta_5 R_t)T_t$. To ensure nonnegative snow depths we insert the index method in an inverse logit function. This means that we assume that the expected portion of the snow (depth) that disappears is given by
\begin{equation}
  \label{eq:3}
  D_{t-1} \mathrm{logit}^{-1}( \beta_3 + (\beta_4 + \beta_5 R_t)T_t )
\end{equation}
Adding the two parts together, given that $D_t > 0$ we assume that the expected snow depth at time $t$, is given by
\begin{equation}
  \label{eq:4}
  E(D_t\, |\, D_t > 0) = e^{\mu} + \underbrace{R_t\, \beta_0\, \mathrm{logit}^{-1}(\beta_1 + \beta_2 T_t)}_{\mathrm{snowfall}} + \underbrace{D_{t-1}\, \mathrm{logit}^{-1}(\beta_3 + (\beta_4 + \beta_5\, R_t)T_t)}_{\mathrm{melting/aging}}
\end{equation}
Since the expectation in \eqref{eq:4} always is larger than zero, we simply use the identity link function when linking the expectation to the gamma distribution. The intercept $e^{\mu}$ typically becomes very small (see Table \ref{tab:1}).

In gamma regression the far most common is to assume that the shape parameter, $k$, is constant such that the coefficient of variation becomes constant
\begin{equation}
  \label{eq:6}
  \frac{SD(D_t)}{E(D_t)} = 1/\sqrt{k} = \mathrm{const}
\end{equation}
where $SD(D_t)$ denotes the standard deviation of $D_t$. Under this parameterization the standard deviation of $D_t$ increases with the expected value of $D_t$. This is not a natural assumption for the modeling snow depth and turns out to give a very poor fit to the data. If the temperature is below zero $^{\circ}$C and it is no precipitation, the changes in snow depth is small, and $D_t$ can be predicted with high precision from $D_{t-1}$ even when $D_{t-1}$ is high. A more natural assumption is that the variance depends on the expected \textit{change} in snow depth. The larger expected changes, the larger uncertainty. To avoid the possibility that the variance is equal to zero, we combined this with the assumption of constant variance resulting in the following model for the variance
\begin{equation}
  \label{eq:21}
  \mathrm{Var}(D_t|D_t>0) = \sigma_1^2 + \sigma_2^2 (E(D_t|D_t>0) - D_{t-1})^2
\end{equation}
Given the expectation and variance, the shape, $k$, and scale, $\theta$, parameter of the gamma distribution can be computed in the usual way
\begin{align}
  \label{eq:13}
  \begin{split}
  k &= \frac{E(D_t\,|\, D_{t}>0)^2}{\mathrm{Var}(D_t\,|\, D_{t}>0)}\\
  \theta &= \frac{\mathrm{Var}(D_t\,|\, D_{t}>0)}{E(D_t\,|\, D_{t}>0)}
  \end{split}
\end{align}
An alternative to the model formulation above is to fit the data with a double generalized linear model, see e.g. the \verb|dglm| package \citep{dglm} in R \citep{R}. We have not tried this.

Next we turn to $P(D_t = 0)$ in \eqref{eq:5} which is modeled by logistic regression. One may estimate this probability using the covariates $R_t$, $T_t$ and $D_{t-1}$. Instead we use $E(D_t\,|\, D_t>0)$ as the only covariate
\begin{equation}
  \label{eq:7}
  P(D_t = 0) = \mathrm{logit}^{-1}(\beta_6 + \beta_7 E(D_t\,|\, D_t>0))
\end{equation}
which turns out to perform well. The intuitive is that higher expected snow depth, given by \eqref{eq:4}, results in lower probability of no snow.

Finally, given the covariates $R_t, T_t, D_{t-1}$ and $R_{t'}, T_{t'}, D_{t'-1}$ for $t' \neq t$, we assume that $D_t$ and $D_{t'}$ are independent. This makes it straight forward to put up the likelihood function for the snow depth observations. Due to the nonlinearities to the covariates and the coupling between the logistic and gamma part of the model, to the best of our knowledge there exists no statistical packages in R \citep{R} or elsewhere to estimate the parameters of the model. Instead we estimate the parameters by implementing a steepest descent optimization algorithm and find the parameters the optimize the likelihood function.

Given the estimated parameters of the model, forecasting of future snow depths can be computed using Monte Carlo simulations. We assume that a reliable weather forecast of $T_t$ and $R_t$ is available for the next few days. The model above can then be used to ``track'' the probability distribution of snow depths into the future as follows. Given the current snow depth $D_t$ and weather forecasts of temperature and precipitation the next day ($T_{t+1}$ and $R_{t+1}$), generate a large set of realizations from the distribution $f(D_{t+1})$ in \eqref{eq:5}. Next, for each sample from $f(D_{t+1})$ and given weather forecasts two days into the future ($T_{t+2}$ and $R_{t+2}$), generate a sample from $f(D_{t+2})$. In such a way we track the snow depth into the future given the weather forecasts of temperature, $T_{t+1}, T_{t+2}, \ldots$, and precipitation, ($R_{t+1}, R_{t+2}, \ldots$.

We tried some extensions to the model as described below, but neither of them improved the model with respect to the Akaike information criterion (AIC).
\begin{itemize}
\item One may expect an unsymmetry for the snowfall inverse logit function and thus we considered the extension $R_t\, \beta_0\, \mathrm{logit}^{-1}(\beta_1 + \beta_2 T_t + \beta_8 T_t^2)$.
\item For the melting part of the model one may argue that if $D_{t-1}$ is large, not the whole snow pack will be exposed for the air temperature and snow melting may go slower. We therefore tested the extension of the melting part of the model $D_{t-1} \mathrm{logit}^{-1}( \beta_3 + (\beta_4 + \beta_5 R_t)T_t + \beta_8 D_{t-1})$.
\item We also conditioned on previous values of snow depth adding the term $\exp{(\beta_8 + \beta_9 D_{t-2})}$ to \eqref{eq:4}.
\end{itemize}

\section{Long term forecasting of snow depth}
\label{sec:long}

Weather forecasts for temperature and precipitation are typically reliable for three to six days into the future. In this section, we build models to forecast snow depth further into the future than this timespan. We consider two different strategies
\begin{itemize}
\item Model 1: When reliable weather forecasts are not available, we use historical observations of precipitation and temperature to build statistical time series models for these variables. The forecasts from these models are further used as input to the model in the previous Section.
\item Model 2: We build a time series model for the snow depth data directly.
\end{itemize}
The strength of model 1 is that the model gives us simultaneous forecasts of temperature, precipitation and snow depth trends.

\subsection{Model 1}
\label{sec:model1}

Let $s(t)$ denote the day during a season for observation time $t$. E.g. if $t$ refers to December 31 for some year, $s(t) = 365$ for ordinary years and 366 for leap years. For long-term forecasts of temperature and precipitation, the seasonal trends will be important. We apply Fourier series, which are able to model complex seasonal patterns with only a few parameters
\begin{equation}
  \label{eq:9}
  h_m(t) = a_0 + \sum_{k=1}^m a_k \sin\left(k\frac{2\pi}{366} s(t)\right) + b_k \cos\left(k\frac{2\pi}{366} s(t)\right)
\end{equation}
Except for the seasonal trend, temperature data fits well to an autoregressive process. Thus, we model the temperature time series using an Autoregressive process with a seasonal trend given by \eqref{eq:9}
\begin{equation}
  \label{eq:10}
  T_t - h_{m_T}(t) = \sum_{j=1}^{p_T} \alpha_j (T_{t-j} - h_{m_T}(t-j)) + \epsilon_t
\end{equation}
where $\epsilon_t \sim N(0, \sigma_T)$ where $N(\mu, \sigma)$ denote a normal distribution with expectation $\mu$ and standard deviation $\sigma$. Further we assume that $\epsilon_t, t=1,2,\ldots,n$ are independent. Standard packages in \citet{R} can be used to fit this model, but instead we found the parameters that maximized the likelihood function using a steepest descent optimization algorithm.

For precipitation we follow the model in \citet{Stern84} with some exceptions that will be explained below. Because of the many days with no precipitation the zero-inflated gamma model in \eqref{eq:5} is suitable also to model precipitation
\begin{equation}
  \label{eq:11}
  f(R_t) = P(R_t = 0)\, \mathrm{I}(R_t = 0) + P(R_t > 0) g(R_t)\, \mathrm{I}(R_t > 0)
\end{equation}
Similar to the model for snow depth and the model in \citet{Stern84} we model $g(R_t)$ with a gamma regression model. We define the expectation as
\begin{equation}
  \label{eq:12}
  E(R_t) = \exp{ \left( h_{m_R}(t) + \sum_{j=1}^{q_R} \gamma_{Rj} R01_{t-j} + \sum_{j=1}^{s_R} \kappa_{Rj} T^j \right) }
\end{equation}
were $R01_t$ is defined such that $R01_t = 1$ if $R_t > 0$ and $R01_t = 0$ if $R_t = 0$. \citet{Stern84} also considers interactions between $R01_{t-j}$ for different values of $j$. We achieve almost as good fit with respect to AIC by instead increasing the value of $q$. The difference in AIC using interactions or not were between 2 and 8 for the data series considered in this paper. For simplicity, we therefore omitted interactions, which made the model easier to interpret. In contrast to \citet{Stern84} we also include the current temperature as a predictor, which results in a substantial improvement of the model. The standard approach of holding the coefficient of variation constant (equation \eqref{eq:6}) resulted in a good fit to the precipitation data.

In \citet{Sloughter07,Moller13} it is suggested to model $R_t^{1/3}$ in stead of $R_t$ with a gamma distribution. Based on goodness of fit analyzes we were not able to show that one of these alternatives resulted in a better fit then the other and decided to model $R_t$ as gamma distributed as shown above.

Similar to the snow depth model, we model $P(R_t = 0)$ with logistic regression. For the snow depth data using $E(D_t\,|\, D_t > 0)$ as the only covariate performed well, but using only $E(R_t\,|\,R_t > 0)$ as a covariate turns out to perform poorly for the precipitation data. A better fit is achieved using the same covariates as above
\begin{equation}
  \label{eq:14}
  P(R_t = 0) = \mathrm{logit}^{-1} \left( h_{m_{R_0}}(t) + \sum_{j=1}^{q_{R_0}} \gamma_{R_0j} R01_{t-j} + \sum_{j=1}^{s_{R_0}} \kappa_{R_0j} T^j \right) 
\end{equation}
Also for this model, we used a steepest descent optimization algorithm to find the parameters that maximized the likelihood function.

Given forecasts of temperature and precipitation using the models above, the model in Section \ref{sec:short} can further be used to forecast snow depth.

\subsection{Model 2}
\label{sec:model2}

While model 1 in the previous section perform long term forecasts of snow depth by first doing long term forecasts of temperature and precipitation, in this section we instead model the snow depth time series directly. The model will be exactly as the model in Section \ref{sec:short} except that the covariates must be changed since precipitation and temperature is unknown. Therefore we change \eqref{eq:4} with
\begin{equation}
  \label{eq:15}
  E(D_t\, |\, D_t>0) = \exp{ \left( h_{m_D}(t) + \sum_{j=1}^{q_D} \gamma_{Dj} D01_{t-j} + \sum_{j=1}^{s_D} \eta_{Dj} D_{t-j} \right) }
\end{equation}
were $D01_t$ is defined such that $D01_t = 1$ if $D_t > 0$ and $D01_t = 0$ if $D_t = 0$. It turned out that using both the covariates $D01_{t-j}$ and $D_{t-j}$ resulted in a better fit than using only $D01_{t-j}$ or only $D_{t-j}$.

\subsection{Monte Carlo procedure}

For the first days in to the future when reliable weather forecast of temperature and precipitation is available, the Monte Carlo method described in Section \ref{sec:short} will be used. Let $\delta$ denote the number of days with reliable weather forecasts. Now suppose that we want to forecast $D_{t+\delta+1}$. After running the Monte Carlo method in Section \ref{sec:short} we have a set of realizations of the time series $D_{t+1}, \ldots, D_{t+\delta}$. Long-term forecasts can then be computed as follows
\begin{itemize}
\item Using model 2, forecasting of $D_{t+\delta+1}$ can be achieved by generating a realization from model 2 conditioned on each of the realizations of the time series $D_{t+1}, \ldots, D_{t+\delta}$. 
\item Using model 1, first a large set of realizations of $T_{t+\delta+1}$ and $R_{t+\delta+1}$ is generated conditioned on the weather forecasts $T_{t+1}, \ldots, T_{t+\delta}$ and $R_{t+1}, \ldots, R_{t+\delta}$. For each realization of $T_{t+\delta+1}$ and $R_{t+\delta+1}$, a realization of $D_{t+\delta+1}$ is generated using the model in Section \ref{sec:short}.
\end{itemize}
The procedures above can be repeated for as long into the future that forecasts of snow depth is needed.

\subsection{Goodness of fit}
\label{sec:gof}

Assume that $X$ is a stochastic variable with a cumulative distribution function $F_X(x)$. It's a well-known fact that $F_x(X) \sim U[0,1]$ where $U[0,1]$ denote a uniform distribution on the $[0,1]$ interval. The procedure can be used for the different models presented above. For the model in Section \ref{sec:short} the cumulative distribution for $D_t$ can be computed as
\begin{align}
  \label{eq:17}
  \begin{split}
    F_{D_t}(d) = &P(D_t \leq d)= \mathrm{logit}^{-1}(\beta_6 + \beta_7 E(D_t\, |\, D_t>0))+\\& + \mathrm{I}(d > 0) (1 - \mathrm{logit}^{-1}(\beta_6 + \beta_7 E(D_t\, |\, D_t>0)))\int_0^{d} g(D) dD
  \end{split}
\end{align}
where $\int_0^{d} g(D) dD$ is the cumulative gamma distribution. Let $d_1, d_2, \ldots, d_n$ denote the real observations of snow depth. If the model fits the data well, we expect the distribution of $F_{D_1}(d_1), F_{D_2}(d_2), \ldots, F_{D_n}(d_n)$ to be close to uniformly distributed. In the computation of $E(D_t\, |\, D_t>0)$, the real observations of temperature, precipitation and snow depth from the previous day are used as input according to \eqref{eq:4}. The same procedure will be used also for the other models above.

\section{Real data example}
\label{sec:example}

In this section, we forecast future snow depths using the models introduced in the previous sections. We downloaded average daily temperature, precipitation and snow depth from three locations in Norway from the web portal \verb|eklima.met.no|. Properties of the three locations are shown in Table \ref{tab:1}.
\begin{table}
  \caption{Properties of meteorological data.}{
  \begin{tabular}{ccc}
    \hline
    Place & Altitude (m) & Timespan with observations \\ \hline
    Oslo (Blindern) & 97 & June 11 1955 $\,-\,$ June 11 2015 \\
    Geilo & 810 & September 1 1966 $\,-\,$ November 30 2006 \\
    Troms\o & 100 & January 1 1955 $\,-\,$ September 1 2015 \\ \hline
  \end{tabular}}
  \label{tab:1}
\end{table}
The locations represent very different climates in Norway. Oslo and Troms\o\, are close to the coast in the southeast and far north of Norway respectively, while Geilo is far inland in the mountainous parts of Norway. 

We now fit the model in Section \ref{sec:short} for each of the three locations using the time series of temperature, precipitation and snow depth. Table \ref{tab:1} shows the estimated parameters for the three locations.
\begin{table}
  \caption{Estimated parameters of the model in Section \ref{sec:short}.}{
  \begin{tabular}{ccccccccccccc}
    \hline
    Place & $\mu$ & $\beta_0$ & $\beta_1$ & $\beta_2$ & $\beta_3$ & $\beta_4$ & $\beta_5$ & $\beta_6$ & $\beta_7$ & $\sigma_1^2$ & $\sigma_2^2$ \\\hline
    Oslo & $-6.92$ & 0.96 & 0.88 & $-$1.76 & 1.99 & $-$0.30 & $-$0.03 &  4.13 & $-$1.97 & 0.63 & 1.79\\
    Geilo & $-5.88$ & 0.72 & 1.74 & $-$1.19 & 2.86 & $-$0.25 & $-$0.05 & 3.61 & $-$0.75 & 1.04 & 2.83\\
    Troms\o & $-4.23$ & 0.89 & 2.02 & $-$0.83 & 2.64 & $-$0.16 & $-$0.04 & 3.38 & $-$0.64 & 0.98 & 8.56\\ \hline
  \end{tabular}}
  \label{tab:1}
\end{table}

\begin{figure}
  \centering
  \includegraphics[width = \textwidth]{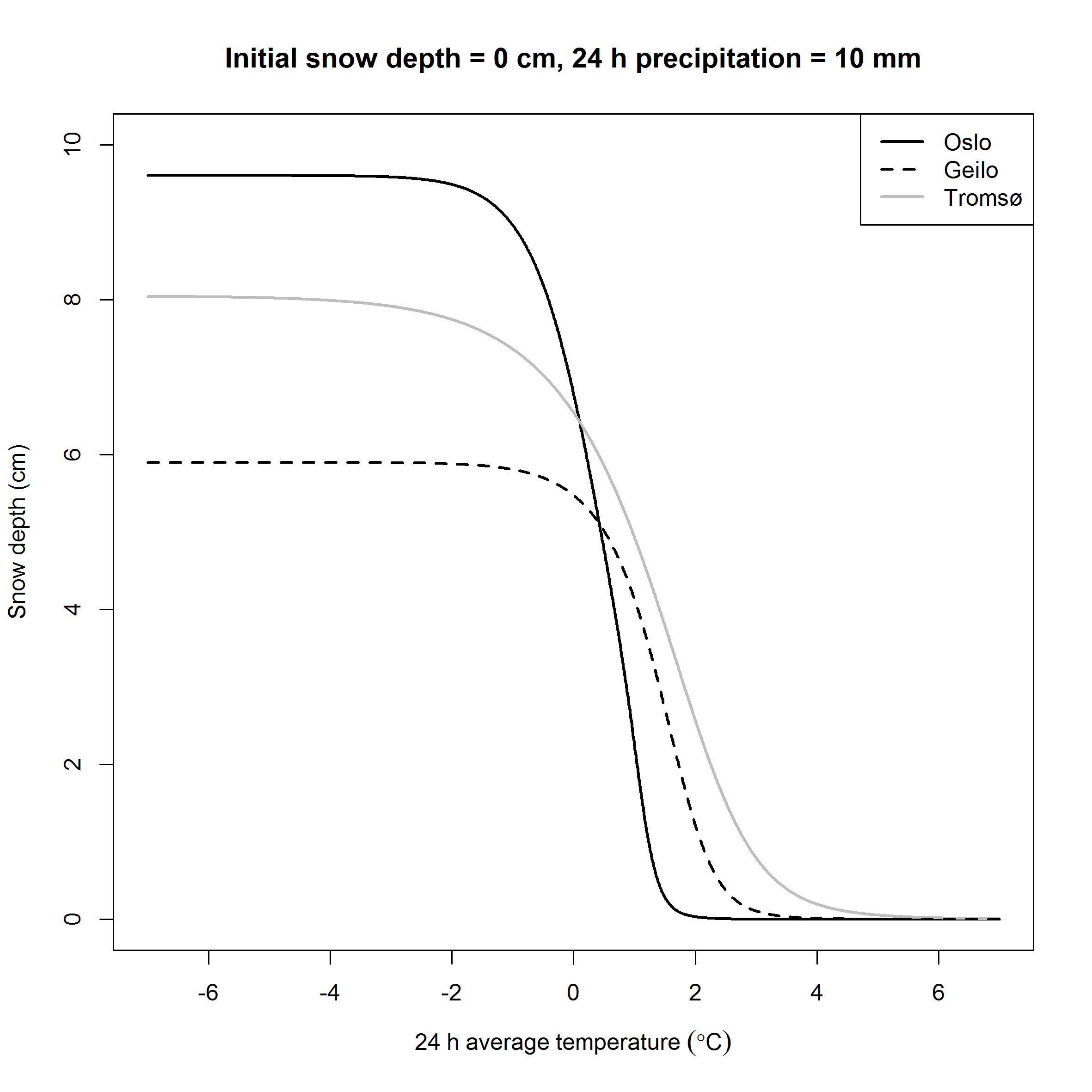}
  \caption{Relation between expected snow depths from 10 mm precipitation for different temperatures.}
  \label{fig:1}
\end{figure}

\begin{figure}
  \centering
  \includegraphics[width = \textwidth]{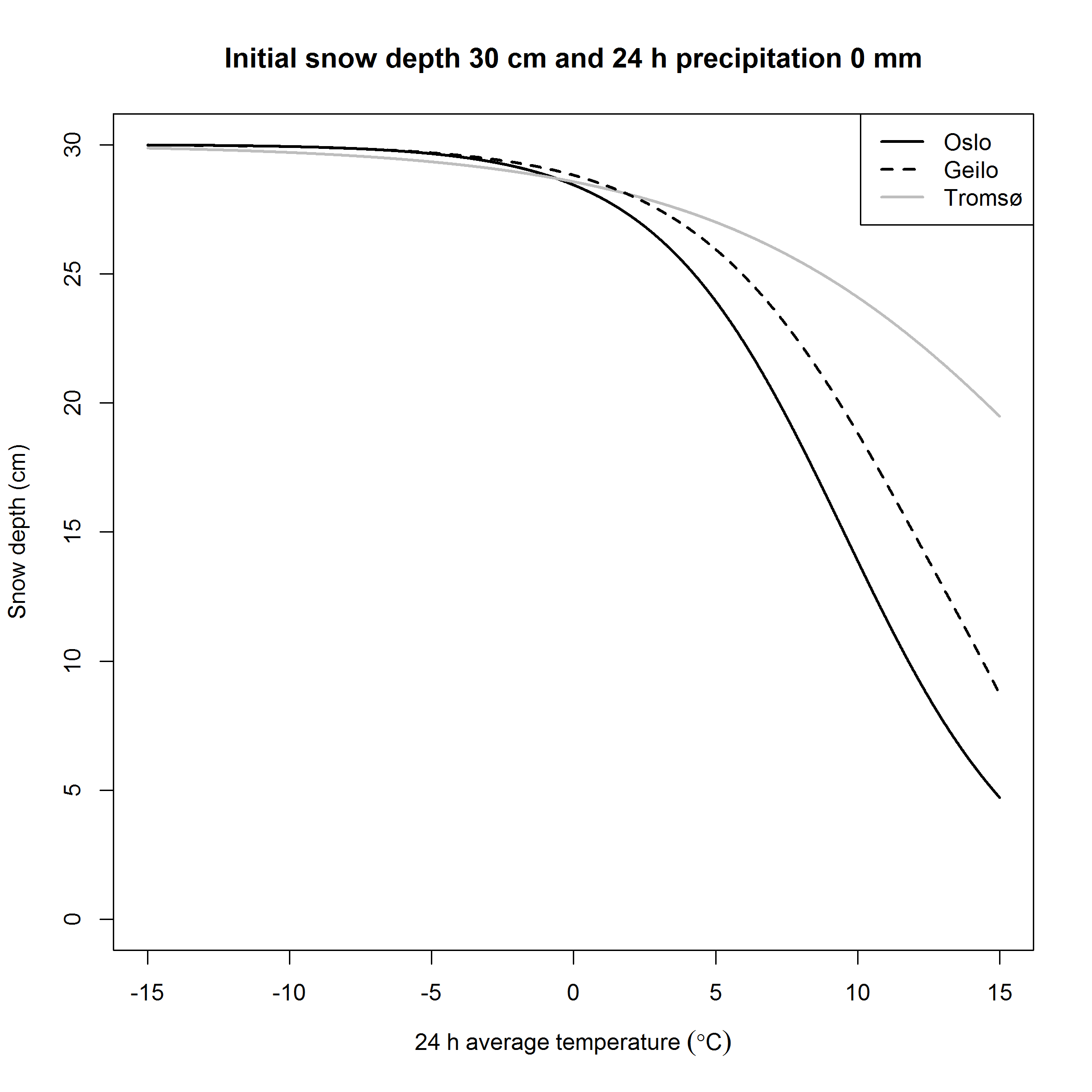}
  \caption{Relation between $E(D_t)$ and temperature when $D_{t-1} = 30$ cm and $R_t = 0$ mm, i.e. the melting/aging process.}
  \label{fig:2}
\end{figure}

\begin{figure}
  \centering
  \includegraphics[width = \textwidth]{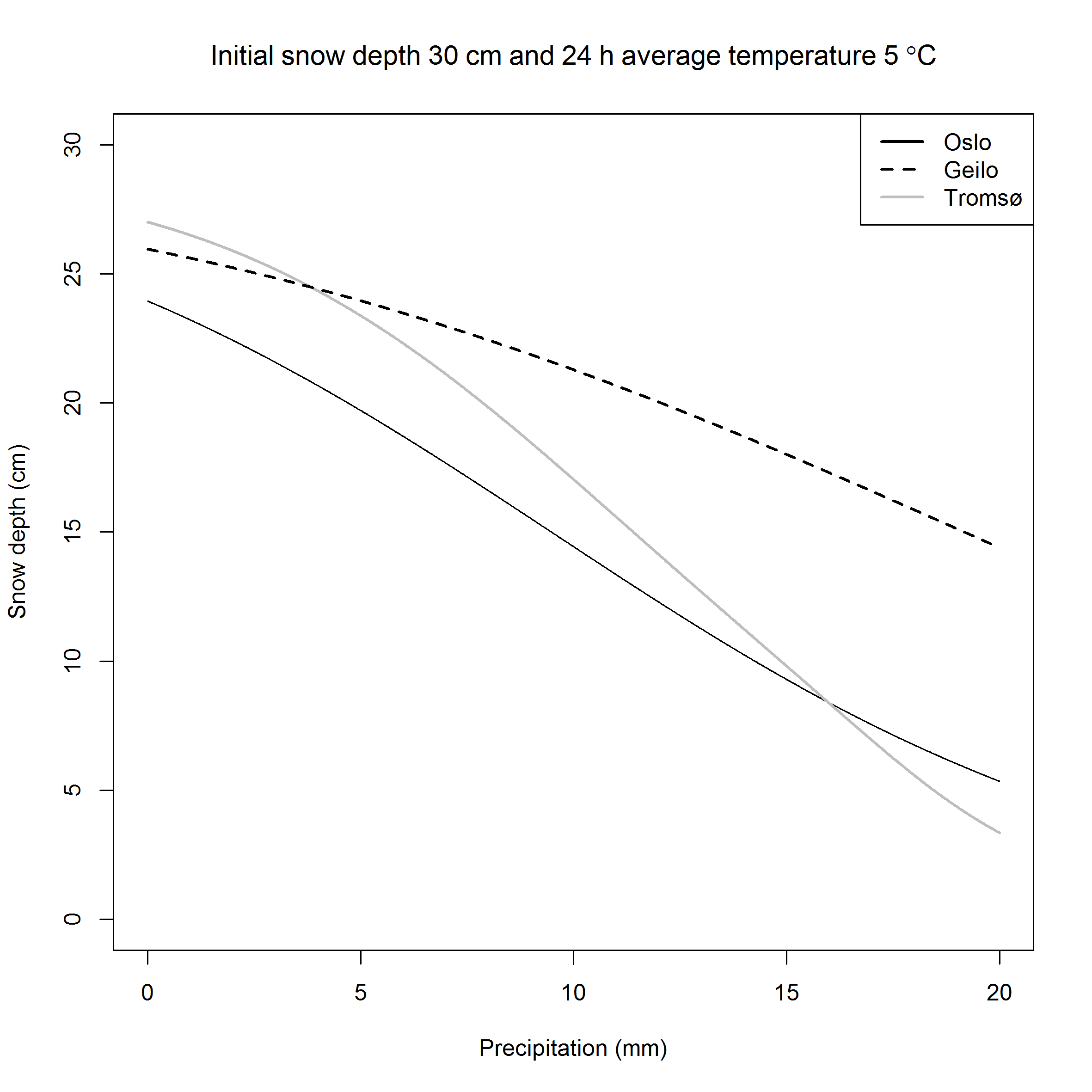}
  \caption{Relation between $E(D_t)$ and precipitation when $D_{t-1} = 30$ cm and $T_t = 0\,^{\circ}$C, i.e. the melting/aging process.}
  \label{fig:3}
\end{figure}

For the fitted models Figures \ref{fig:1}, \ref{fig:2} and \ref{fig:3} show curves for $E(D_t)$ for different values of $T_t$, $R_t$ and $D_{t-1}$. $E(D_t)$ are computed using the law of total expectation
\begin{align}
  \label{eq:16}
  \begin{split}
    E(D_t) &= E(D_t\, |\, D_t = 0)P(D_t = 0) + E(D_t\, | D_t > 0)P(D_t > 0) \\
           &= E(D_t\, | D_t > 0)P(D_t > 0)
  \end{split}
\end{align}
where $E(D_t\, | D_t > 0)$ and $P(D_t > 0)$ are computed using \eqref{eq:4} and \eqref{eq:7}, respectively.

Figure \ref{fig:1} snows expected snow depth from 10 mm precipitation for different temperatures. We set $D_{t-1} = 0$ so that we only look at the snowfall part of the model, recall \eqref{eq:4}.
As expected the snow depth decreases rapidly around $0\, ^{\circ}$C and the curves are in accordance with earlier research on such curves \citep{Wen13,Kienzle08}. Due to differences in climate the curves varies with location in Norway and such differences is also observed in earlier research \citep{Wen13,Kienzle08}.

Figure \ref{fig:2} shows $E(D_t)$ for different temperatures when assuming that $R_t = 0$ mm and $D_{t-1} = 30$ cm. Since $R_t$ is set to zero, this figure shows the melting/aging part of the model.
Figure \ref{fig:3} shows $E(D_t)$ for different amounts of precipitation when $T_t = 5\,^{\circ}$C and $D_{t-1} = 30$ cm. This figure shows the effect of precipitation on the reduction of snow depth. We see that rain has a strong impact on the reduction of snow depth. We also observe that the curves seem to be quite linear which is in accordance with the index model \citep{Scherrer13}.
The Figures \ref{fig:1} $-$ \ref{fig:3} show that the model captures the main effects of snow accumulation, aging and melting.

To perform long therm forecasts of snow depth, we fitted the time series of temperature, precipitation and snow depth using the models in Section \ref{sec:long}. In each of these models the number of covariates where chosen based on AIC in a forward stepwise regression procedure. E.g. for temperature, first $m_T$ is increased to one, then $p_T$ to one, then $m_T$ to two and so on. The results are shown i Table \ref{tab:3}.
\begin{table}
  \caption{Number of parameter included in the precipitation, temperature and snow depth models.}{
  \begin{tabular}{cccccccccccc}
    \hline
      & \multicolumn{2}{c}{Temperature} & \multicolumn{6}{c}{Precipitation} & \multicolumn{3}{c}{Snow depth} \\
    Place & $m_T$ & $p_T$ & $m_R$ & $q_R$ & $s_R$ & $m_{R_0}$ & $q_{R_0}$ & $s_{R_0}$ & $m_D$ & $q_D$ & $s_D$ \\ \hline
    Oslo            &2&3&3&5&4&3&5&4&3&1&5\\
    Geilo           &2&3&2&3&3&3&5&3&3&1&3 \\
    Troms\o         &2&4&3&6&3&2&6&3&3&1&6\\ \hline
  \end{tabular}}
  \label{tab:3}
\end{table}

\subsection{Goodness of fit}

\begin{figure}
  \centering
  \begin{tabular}{cc}
    \includegraphics[width = 0.5\textwidth]{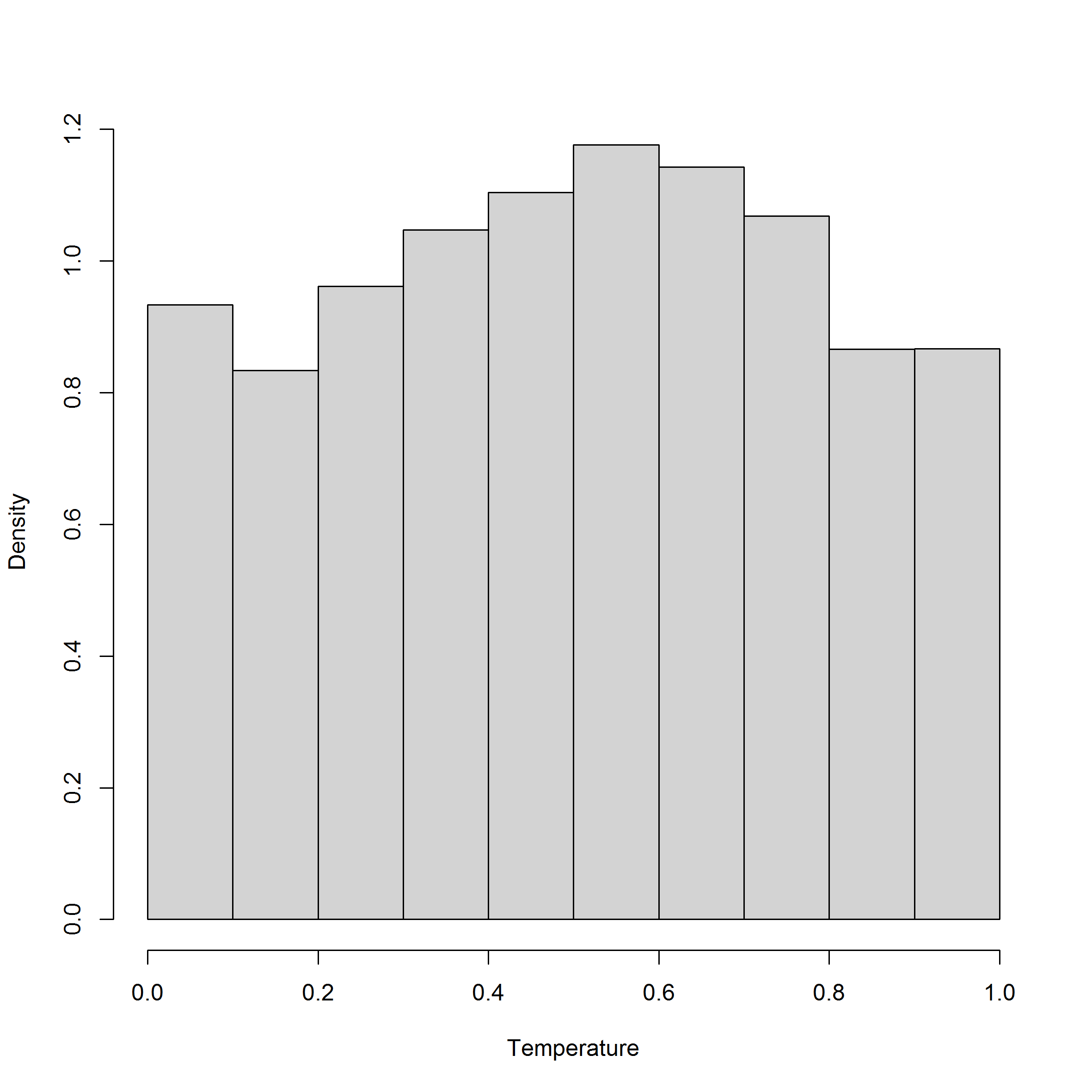} & \includegraphics[width = 0.5\textwidth]{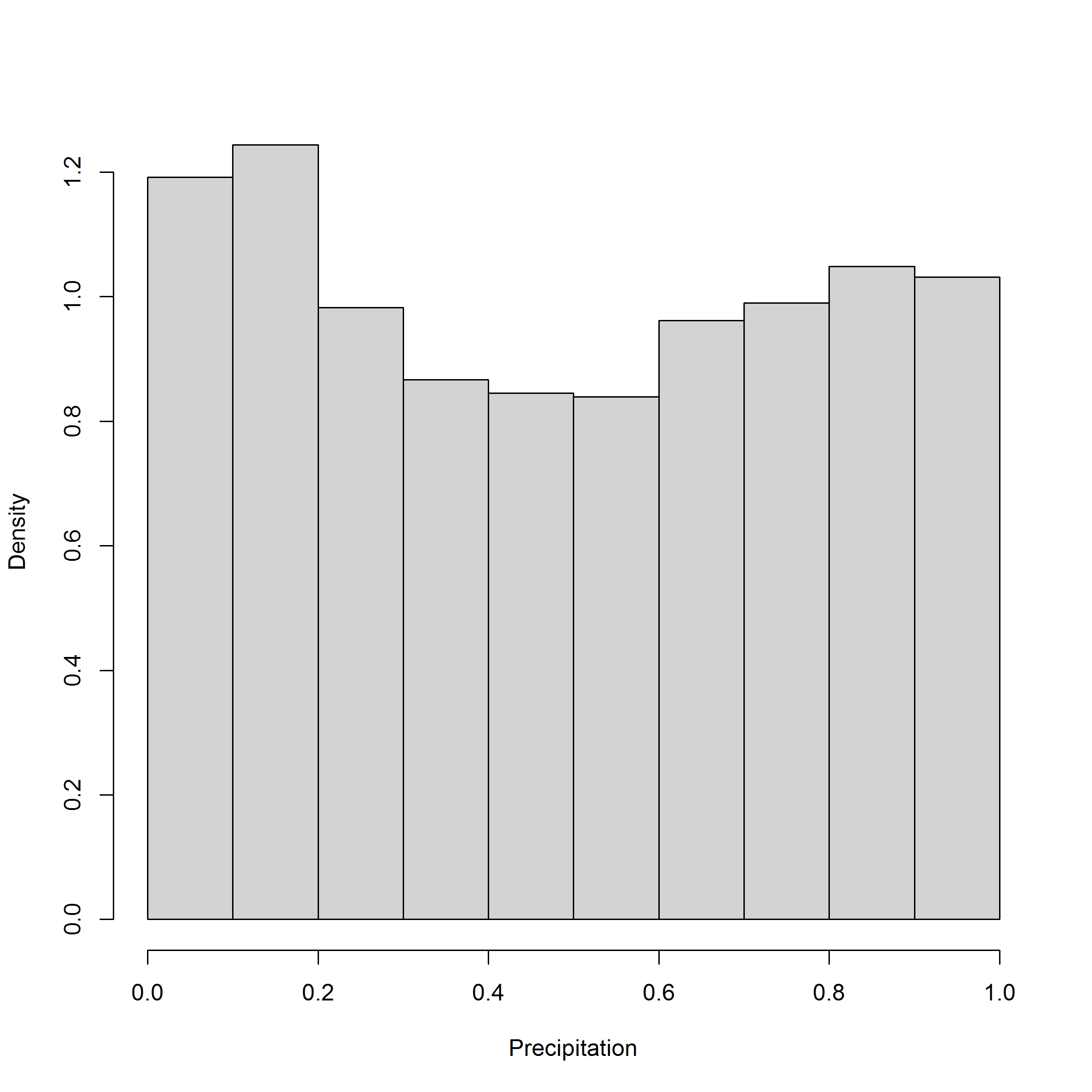}\\
    \includegraphics[width = 0.5\textwidth]{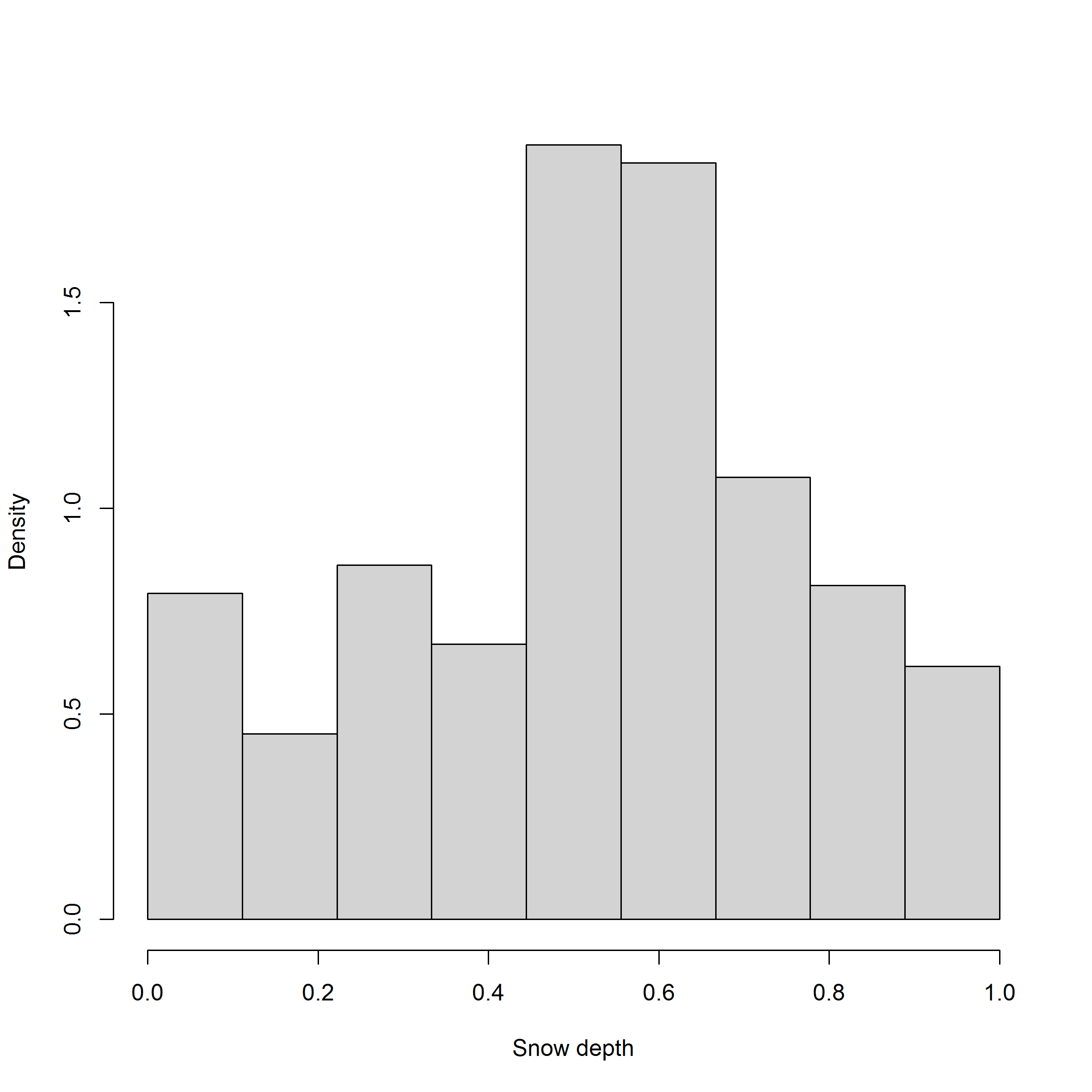} & \includegraphics[width = 0.5\textwidth]{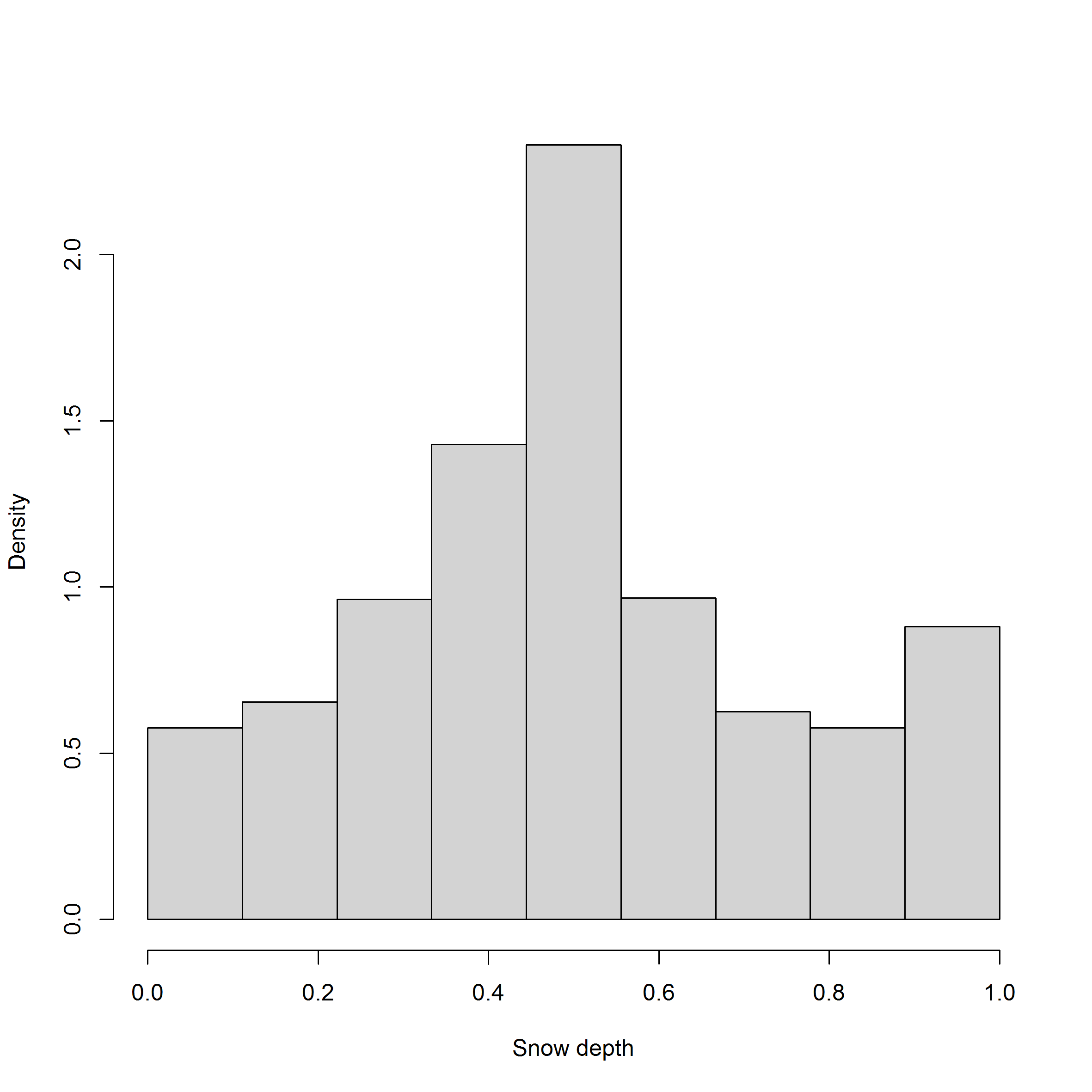}\\
  \end{tabular}
  \caption{Goodness of fit histograms for Oslo. Upper row from left to right shows temperature and precipitation, respectively. The bottom row from left to right show goodness of fit histograms for the model in Section \ref{sec:short} and model 2, respectively.}
  \label{fig:7}
\end{figure}

Figure \ref{fig:7} shows goodness of fit histograms of the models in Sections \ref{sec:short} and \ref{sec:long} using the approach described in Section \ref{sec:gof}. The upper row in Figure \ref{fig:7} shows from left to right goodness of fit histograms for temperature and precipitation, respectively, while the bottom row shows from left to right goodness of fit histograms for the model in Section \ref{sec:short} and model 2, respectively. The goodness of fit histograms for temperature and precipitation are based on observation for the whole year while for the histograms for the snow depth models are based on the winter months December, January and February. All the histograms are for Oslo, but the histograms for Geilo and Troms\o\, were similar.
We see that the histograms for temperature and precipitation looks fairly uniformly distributed. The two models for snow depth (bottom row) show an overrepresentation of values around 0.5. When the temperature is below $0 ^{\circ}$C and it is no precipitation, the changes in snow depth is minimal and the model overestimates the variance in this cases resulting in to many values around 0.5 in the goodness of fit histograms. During winter, these weather conditions is of course very common. We can of course reduce the value of $\sigma_1^2$ in \eqref{eq:21} to reduce the variance when the expected change in snow depth is small, but that will result in other negative consequences for the model.

\subsection{Forecast performance}

We now inspect the forecast performance of the snow depth models. Forecasts are performed using the Monte Carlo procedures described in Sections \ref{sec:short} and \ref{sec:long}. In each time step we forecast using the mean value of the Monte Carlo samples. Classification error is measured by the average difference in absolute value between the observed and forecasted snow depth. Classification is performed for the winter months December, January and February in a cross validation procedure. All data except for one year from July 1 to June 30 the next year is used to fit the models, and further used to forecast snow depths for December, January and February for the year of data not included in the model fitting. The procedure is repeated for each year. We consider three different cases where we assume that reliable weather forecasts of temperature and precipitation are available for zero, five and ten days into the future. For the days were we assume that reliable weather forecasts are available, the real observed values of temperature and precipitation are used, i.e. we assume ``perfect'' weather forecasts. For comparison we also consider a version of model 1 and 2, where all covariates except the periodic covariates are set to zero, i.e. all the variables $p_T$, $q_R$, $s_R$, $q_{R_0}$, $s_{R_0}$, $q_D$, $s_D$ are set to zero which means that only $m_T$, $m_R$, $m_{R_0}$ and $m_D$ can larger than zero. In addition, we assume that the number of days with reliable weather forecasts of precipitation and temperature are zero. In other words, these versions of model 1 and 2 do not take advantage of previous observations for the given season or weather forecasts and forecast only based on seasonal properties. We expect that the further we forecast into the future, the less will the usefulness of previous observations for the given season or weather forecasts be.

\begin{figure}
  \centering
  \includegraphics[width = \textwidth]{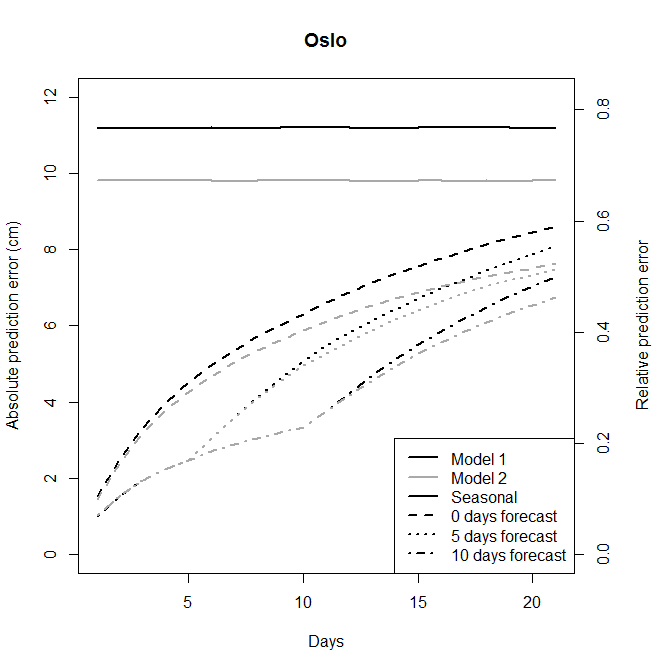}
  \caption{Forecast performance of snow depth for Oslo. The gray and black curves show results for model 1 and model 2, respectively. The dashed, dotted and dash-dotted curves show forecasts when we assume that reliable forecasts of temperature and precipitation are available for zero, five and ten days, respectively. The solid curves show classification performance where only the periodic covariates are included. The left vertical axis show average classification error in absolute value, while the right vertical axis show the classification error normalized with the average snow depth.}
  \label{fig:4}
\end{figure}

\begin{figure}
  \centering
  \includegraphics[width = \textwidth]{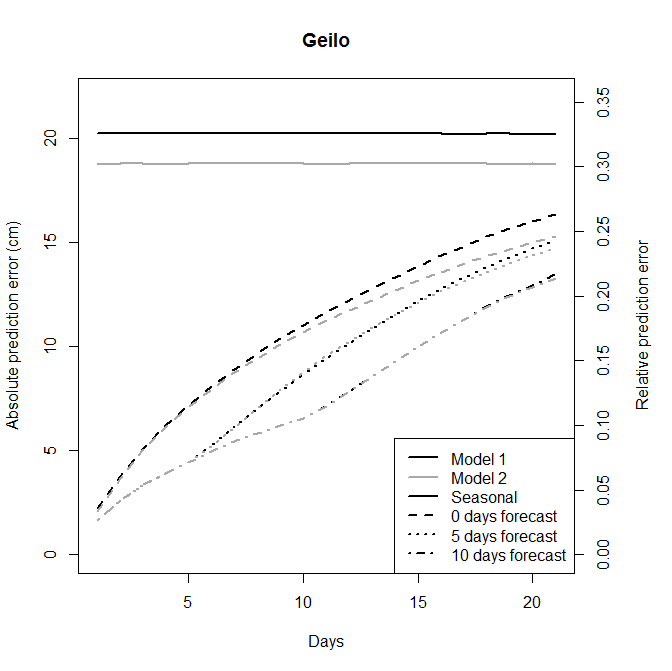}
  \caption{Forecast performance of snow depth for Geilo. The gray and black curves show results for model 1 and model 2, respectively. The dashed, dotted and dash-dotted curves show forecasts when we assume that reliable forecasts of temperature and precipitation are available for zero, five and ten days, respectively. The solid curves show classification performance where only the periodic covariates are included. The left vertical axis show average classification error in absolute value, while the right vertical axis show the classification error normalized with the average snow depth.}
  \label{fig:5}
\end{figure}

\begin{figure}
  \centering
  \includegraphics[width = \textwidth]{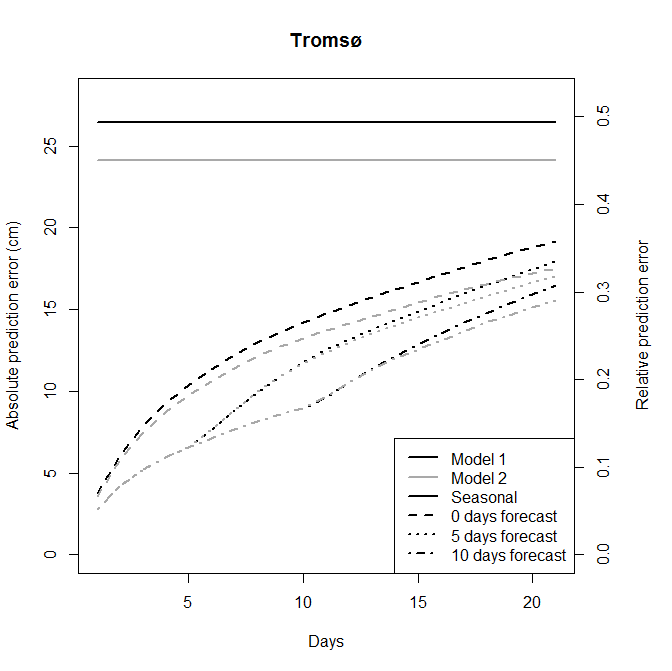}
  \caption{Forecast performance of snow depth for Troms\o. The gray and black curves show results for model 1 and model 2, respectively. The dashed, dotted and dash-dotted curves show forecasts when we assume that reliable forecasts of temperature and precipitation are available for zero, five and ten days, respectively. The solid curves show classification performance where only the periodic covariates are included. The left vertical axis show average classification error in absolute value, while the right vertical axis show the classification error normalized with the average snow depth.}
  \label{fig:6}
\end{figure}

The results for Oslo, Geilo and Troms\o\, are shown in Figures \ref{fig:4} $-$ \ref{fig:6}. 
The gray and black curves show results for model 1 and model 2, respectively. The dashed, dotted and dash-dotted curves show forecasts when we assume that reliable weather forecasts of temperature and precipitation are available for zero, five and ten days, respectively. For the days when reliable weather forecasts are available, the model in Section \ref{sec:short} is used. The solid curves show classification performance where only the periodic covariates are included as explained above. The left vertical axis show average classification error in absolute value, while the right vertical axis show the classification error normalized with the average snow depth for the months of December, January and February.

We see that model 2 perform better then model 1. Further we see that given reliable weather forecasts the classification error is about half compared to not having reliable weather forecasts. We also observe that useful forecasts is possible long in to the future. Forecasting three weeks into the future and given five days of reliable weather forecasts the classification error is a little over half of the classification error using only the periodic covariates. In comparison, weather forecasts of temperature and precipitation are typically completely dominated by the seasonal trends after only a few days. Forecast error is lower for Oslo compared to Geilo and Troms\o\,, but relative to the average snow depth forecast error for Oslo is higher than for Geilo and Troms\o.

\begin{figure}
  \centering
  \includegraphics[width = \textwidth]{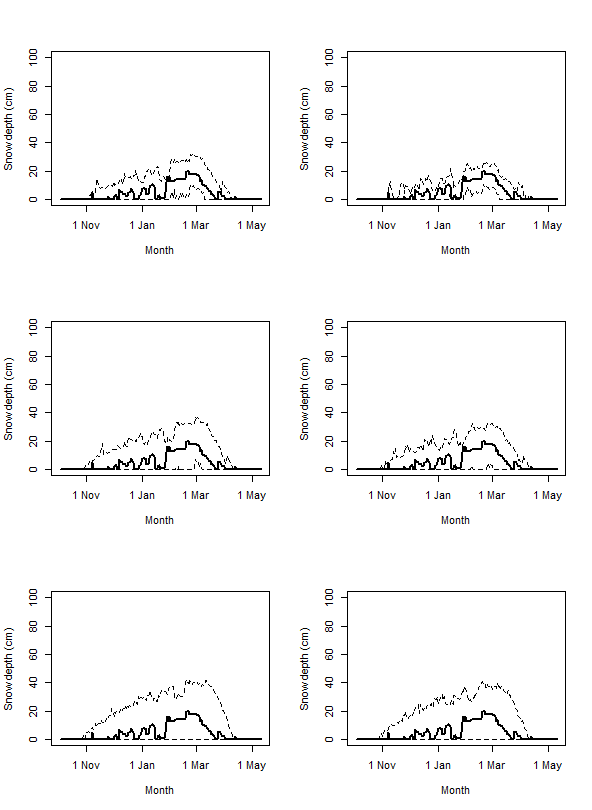}
  \caption{Forecasts of snow depth for Oslo using model 2. The first, second and third row show forecasts for five, ten and three weeks into the future. In the left and the right column, we assume that zero and five days of reliable weather forecasts of temperature and precipitation are available, respectively. The solid curve shows the real snow depth data, while the dashed curves show 5\% and 95\% quantiles of the forecast distribution.}
  \label{fig:13}
\end{figure}

\begin{figure}
  \centering
  \includegraphics[width = \textwidth]{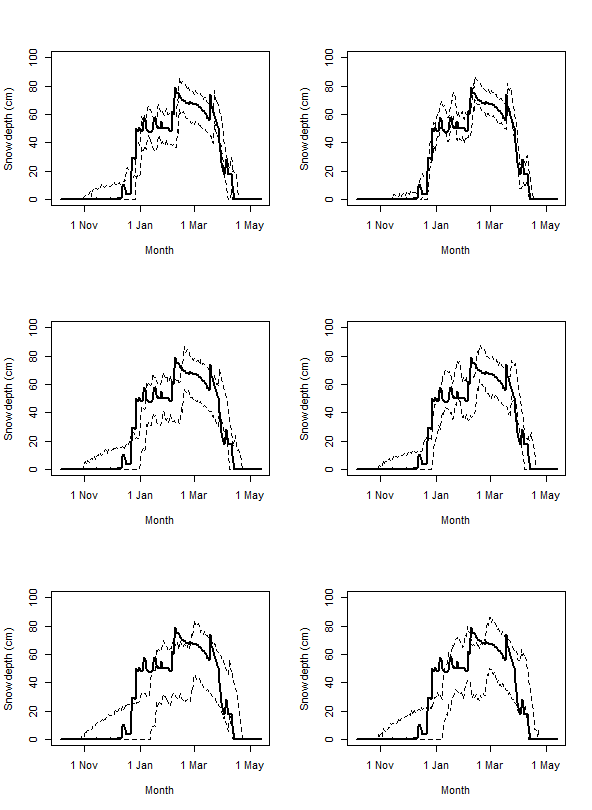}
  \caption{Forecasts of snow depth for Oslo using model 2. The first, second and third row show forecasts for five, ten and three weeks into the future. In the left and the right column, we assume that zero and five days of reliable weather forecasts of temperature and precipitation are available, respectively. The solid curve shows the real snow depth data, while the dashed curves show 5\% and 95\% quantiles of the forecast distribution.}
  \label{fig:8}
\end{figure}

\begin{figure}
  \centering
  \includegraphics[width = \textwidth]{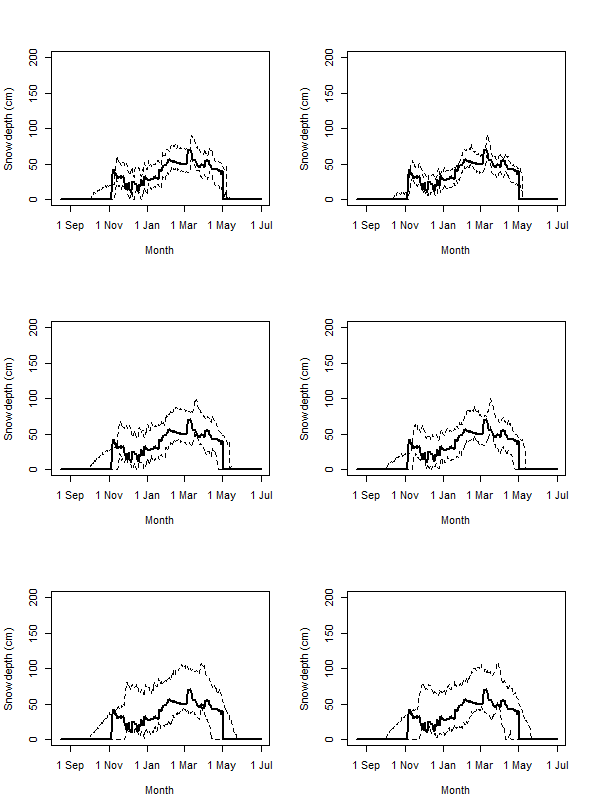}
  \caption{Forecasts of snow depth for Geilo using model 2. The first, second and third row show forecasts for five, ten and three weeks into the future. In the left and the right column, we assume that zero and five days of reliable weather forecasts of temperature and precipitation are available, respectively. The solid curve shows the real snow depth data, while the dashed curves show 5\% and 95\% quantiles of the forecast distribution.}
  \label{fig:9}
\end{figure}

\begin{figure}
  \centering
  \includegraphics[width = \textwidth]{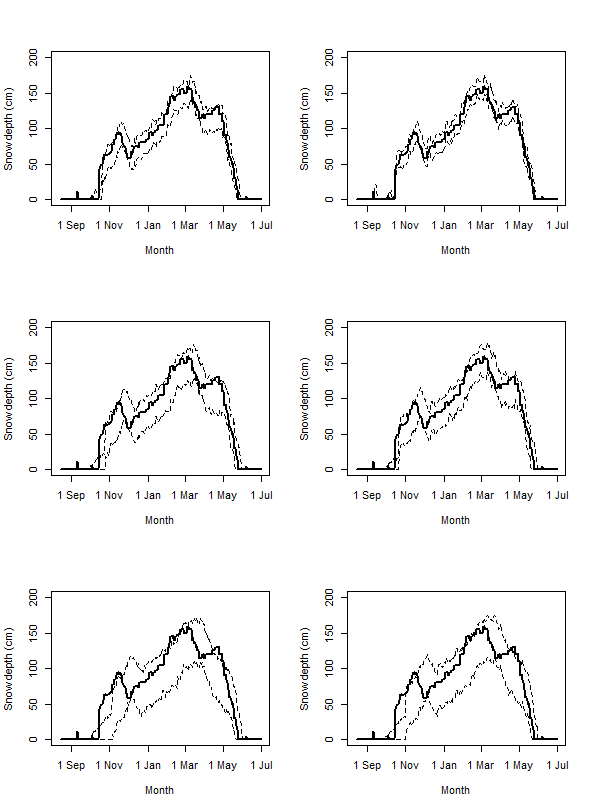}
  \caption{Forecasts of snow depth for Geilo using model 2. The first, second and third row show forecasts for five, ten and three weeks into the future. In the left and the right column, we assume that zero and five days of reliable weather forecasts of temperature and precipitation are available, respectively. The solid curve shows the real snow depth data, while the dashed curves show 5\% and 95\% quantiles of the forecast distribution.}
  \label{fig:10}
\end{figure}

\begin{figure}
  \centering
  \includegraphics[width = \textwidth]{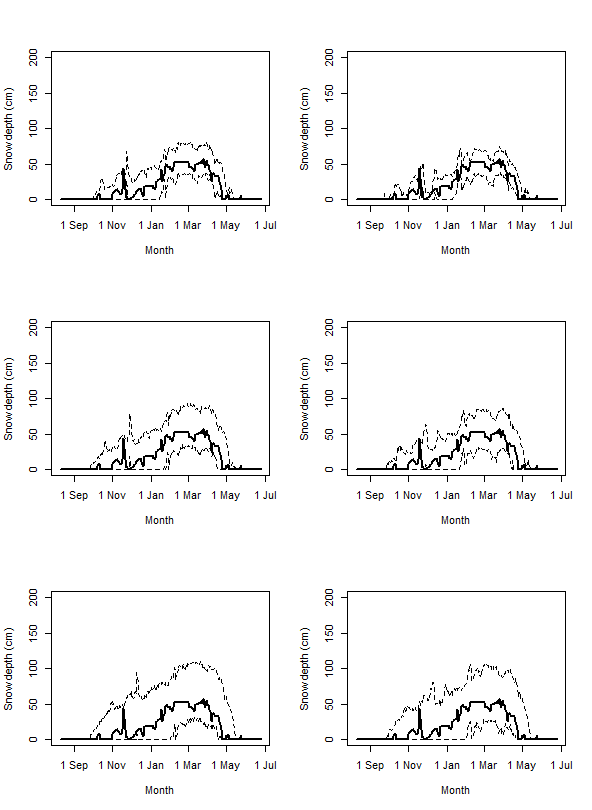}
  \caption{Forecasts of snow depth for Troms\o\, using model 2. The first, second and third row show forecasts for five, ten and three weeks into the future. In the left and the right column, we assume that zero and five days of reliable weather forecasts of temperature and precipitation are available, respectively. The solid curve shows the real snow depth data, while the dashed curves show 5\% and 95\% quantiles of the forecast distribution.}
  \label{fig:11}
\end{figure}

\begin{figure}
  \centering
  \includegraphics[width = \textwidth]{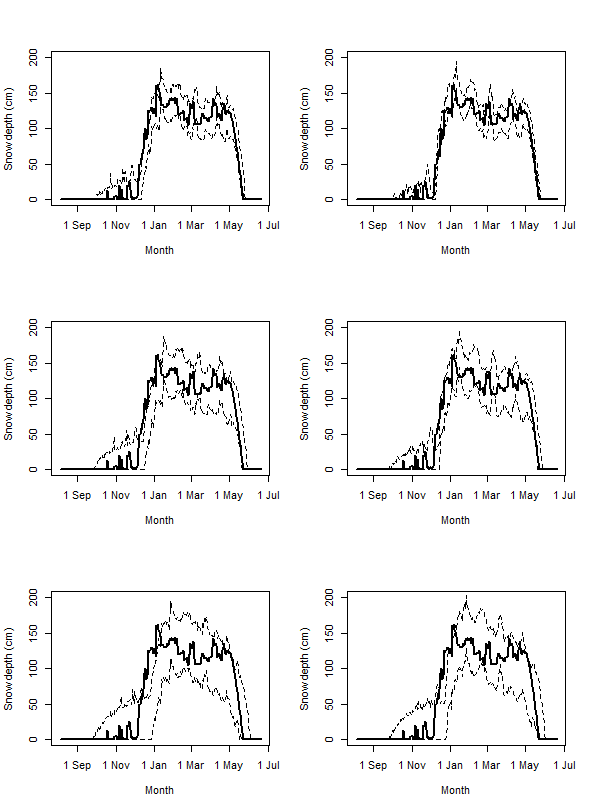}
  \caption{Forecasts of snow depth for Troms\o\, using model 2. The first, second and third row show forecasts for five, ten and three weeks into the future. In the left and the right column, we assume that zero and five days of reliable weather forecasts of temperature and precipitation are available, respectively. The solid curve shows the real snow depth data, while the dashed curves show 5\% and 95\% quantiles of the forecast distribution.}
  \label{fig:12}
\end{figure}

Figures \ref{fig:13} and \ref{fig:8} show forecasts for future snow depths in Oslo using model 2 for a season with little and much snow, respectively. The first, second and third row show forecasts for five, ten and three weeks into the future. In the left and the right column, we assume that zero and five days of reliable weather forecasts of temperature and precipitation are available, respectively. The solid curve shows the real snow depth data, while the dashed curves show 5\% and 95\% quantiles of the forecast distribution.
Figure \ref{fig:9} $-$ \ref{fig:12} show the same for Geilo og Troms\o.
We see that given reliable weather forecasts more precise snow depth forecasts can be achieved. Comparing the bottom rows of Figures \ref{fig:13} and \ref{fig:8}, \ref{fig:9} and \ref{fig:10} and \ref{fig:11} and \ref{fig:12}, we see that the forecasts for three weeks into the future is quite different for seasons with little and much snow which shows that reliable long terms forecasts of snow depth are possible which are more precise than just using seasonal trends.

\section{Closing remarks}
\label{sec:closrem}

This paper presents a first attempt to build statistical models for short and long term forecasts of snow depth. The results show that it is possible to do useful forecasts of snow depth long into the future.
Further we found that model 2 (Section \ref{sec:model2}) perform better then model 1 (Section \ref{sec:model2}), but the advantage of model 1 compared to model 2 is that long term simultaneous scenarios of temperature, precipitation and snow depth is computed. This can be useful in for many applications. E.g. with respect to road safety the risk of slippery roads is especially high when the snow depth is above zero cm and at the same time the temperature is below zero $^{\circ}$C.

Several extensions to the suggested models are possible. Including other covariates like solar radiation, humidity, wind and the age of the snowpack may improve the forecasts \citep{Roebber03,Kuusisto84}. Models that better separate sinking/aging from melting may be achieved by including the water content in the snow as a hidden layer in the model. Light snow tend to sink faster than denser snow and is not possible to separate in the model presented in this paper. Figures \ref{fig:1} $-$ \ref{fig:3} reveal that the properties of snowfall and aging/melting varies at different locations. A natural extension is thus to include the models in this paper as part of a spatio-temporal model.

\bibliography{bibl}

\end{document}